\documentclass[9.5pt, final, journal, letterpaper, oneside]{IEEEtran}

\usepackage{etoolbox}
\usepackage{epsfig, amsmath, amssymb}
\usepackage{algorithm}
\usepackage{algpseudocode}
\usepackage{epstopdf}
\usepackage{balance}
\usepackage{color}
\usepackage{url}
\usepackage{textcomp}
\usepackage{caption}
\usepackage{graphicx}
\usepackage{subcaption}

\usepackage{lipsum}

\definecolor{red}{rgb}{1,0,0}
\definecolor{blue}{rgb}{0,0,1}

\apptocmd{\thebibliography}{\renewcommand{\sc}{}}{}{}

\newcommand{\eg}{{\it e.g.}, }

\newcommand{\etal}{{\it et~al.} }

\title{Toward a Secure and Decentralized Blockchain-based Ride-Hailing Platform for Autonomous Vehicles}

\author{
Ryan Shivers\IEEEauthorrefmark{1}, Mohammad Ashiqur Rahman\IEEEauthorrefmark{2}, and Hossain Shahriar\IEEEauthorrefmark{3}\\

\IEEEauthorrefmark{1}Department of Computer Science, Tennessee Tech University

\IEEEauthorrefmark{2}Department of Electrical and Computer Engineering, Florida International University

\IEEEauthorrefmark{3}Department of Computer Science, Kennesaw State University

Emails: rmshivers42@students.tntech.edu, marahman@fiu.edu, hshahria@kennesaw.edu
}

\begin{document}

\maketitle
\begin{abstract}
Ride-hailing and ride-sharing applications have recently gained in popularity as a convenient alternative to traditional modes of travel.  Current research into autonomous vehicles is accelerating rapidly and will soon become a critical component of a ride-hailing platform's architecture.  Implementing an autonomous vehicle ride-hailing platform proves a difficult challenge due to the centralized nature of traditional ride-hailing architectures. In a traditional ride-hailing environment the drivers operate their own personal vehicles so it follows that a fleet of autonomous vehicles would be required for a centralized ride-hailing platform to succeed. Decentralization of the ride-hailing platform would remove a road block along the way to an autonomous vehicle ride-hailing platform by allowing owners of autonomous vehicles to add their vehicle to a community driven fleet when not in use.  Blockchain technology is an attractive choice for this decentralized architecture due to its immutability and fault tolerance.  This paper proposes a framework for developing a decentralized ride-hailing architecture implemented on the Hyperledger Fabric blockchain platform. The implementation is evaluated using a static analysis tool and performing a performance analysis under heavy network load.
\end{abstract}

\begin{IEEEkeywords}
Blockchain; Hyperledger Fabric; Ride-Hailing; Ride-Sharing; Information Security;
\end{IEEEkeywords}

%%%%%%%%%%%%%%%%%%%%%
\section{Introduction}
\label{Sec:Introduction}

Ride-sharing services fill empty seats in cars with people who are traveling near the same destination as a driver. This concept of ride-sharing has evolved since its inception to a large-scale market as mass appeal has skyrocketed its profitability. The ride-sharing / ride-hailing marketplace has been rapidly expanding since the launch of companies such as Uber and Lyft that offer a platform for cooperation between riders and drivers through mobile applications. Goldman Sachs has predicted that the ride-sharing market revenue will be worth approximately 285 billion dollars by the year 2030 \cite{burgstaller2017rethinking}. This assumes adoption of self-driving car technology in the ride-hailing / ride-sharing market continues to advance at the rate predicted by Goldman Sachs analysts.

Autonomous Vehicles (AVs) collect information about the current state of the environment around them using sensors (\eg cameras, lasers, and electromagnetic field detectors) and feed the information into traditional Artificial Neural Networks to make decisions about how to operate the vehicle while on the road. The first system that operated in this manner was the Autonomous Land Vehicle in a Neural~ Network~(ALVINN)~\cite{alvinn}. The blockchain implementation proposed in this paper addresses the network structure and communications of participants and would not affect the Neural Network style operation of the AVs. Development in the AV field is very promising and is likely to alter our current and future transportation infrastructure. Ride-hailing and ride-sharing applications stand to benefit from utilizing new technologies as they would reduce both operating costs and the safety of the passengers in their network as AVs have been shown to operate at a higher safety standard than a human~driver.

Another disruptive technology that has been developed in the recent years is blockchain. Blockchain was first introduced in 2008 when Satoshi Nakamoto published \cite{bitcoin} which described a peer-to-peer electronic cash system known as Bitcoin. Since then, this concept of peer-to-peer cash systems has been labeled as cryptocurrency and many variations of Mr. Nakamoto's original system have been developed. Bitcoin had a specific purpose as an electronic cash system. However, its underlying technology showed promise for use outside the original goal. The underlying technology of Bitcoin is known as blockchain technology and works as a distributed append only ledger, where all information within the system is stored and accessed by connected peers. There have been several different implementations of this underlying system but most are intended as cryptocurrencies. Lin \etal \cite{lin2017survey} outline the blockchain architecture, different types of blockchains, and security issues and challenges that accompany this technology. One implementation that intends to serve as a platform for private business blockchains is Hyperledger Fabric \cite{hlf}. Blockchain technology could prove useful to ride-hailing and AVs by providing a peer-to-peer infrastructure that can be managed independently from a third party.

\noindent\textbf{Contributions:} Our work intends to utilize the more specialized blockchain platform Hyperledger Fabric to create a decentralized ride-hailing framework where the infrastructure would be maintained by collections of drivers working together. This framework would be beneficial to the development and adoption of a ride-hailing platform specifically geared towards AVs because by nature blockchain technology creates trust between multiple non-trusting entities. ``Drivers" are the owners of the AVs and would utilize this platform to form a network of independent drivers that would function as if created and maintained by a centralized source with all of the benefits of centralization. Drivers would have increased flexibility regarding the price given for a ride and would have no percentage of profit to pay to an overarching entity. Trust between participants in the network is provided through the chaincode protocol proposed in this work. Through the protocol, information is provided only to relevant users and enforced through implicit access control implemented in the chaincode functions. The data security provided by the chaincode functions allows for differing client applications to participate in the network without fear of data theft or tampering. This security is provided through the inherit design of the transaction protocol proposed in this work and the designed underlying blockchain structure where data is stored.  Transactions made using the proposed protocol do not allow for specification of requesting user as the certificate of the requesting user is passed implicitly and used in the transaction.  Additional security is provided through design of chaincode permissions provided by Hyperledger Fabric and traditional encryption of packets.  The platform is shown to operate reliably in high network traffic situations and with different configurations of nodes in the performance evaluation of this work. The implementation in this paper is transportation system agnostic in that it is not specific to AV ride-hailing and could be used as a decentralized ride-hailing platform for standard human driver ride-hailing. 

The rest of this paper will be organized as follows: Section~\ref{Sec:Background} provides background information such as descriptions of related technologies and challenges that we overcome during the development of the framework described in this paper, Section~\ref{Sec:Related-Work} describes research that has already been done in related fields, Section~\ref{Sec:Framework} describes the proposed framework, and Section~\ref{Sec:Implementation} details the implementation of our framework. Section~\ref{Sec:Case Study} discusses a case study involving a simulation using real world locations, Section~\ref{Sec:Evaluation} evaluates the security and load resiliency of our implementation, and Section~\ref{Sec:Conclusion} concludes the paper and discusses future work.

%%%%%%%%%%%%%%%%%%%%%

%%%%%%%%%%%%%%%%%%%%%%%%%%%%%%%%%%%%%%%%%%%%%%%%%%%%%%%%%%%%%%%%%%%%%%%%%%%%%
\section{Research Background and Motivation}
\label{Sec:Background}

This section will describe challenges that were anticipated with this work and our motivation for creating and implementing this framework. A brief background of blockchain technology, autonomous vehicles, and ride-sharing services is provided as well. 

\begin{figure}[t]
	\begin{center}
	\includegraphics[width=0.8\columnwidth]{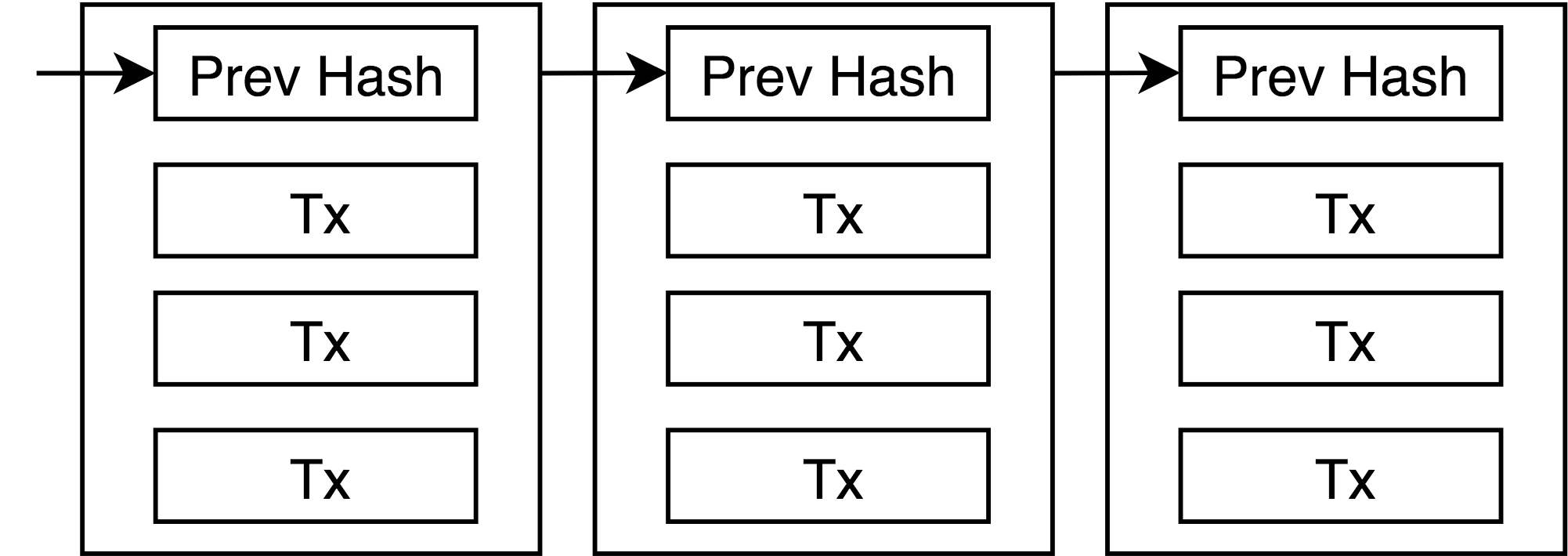}
	\caption{Simplified Chain of Hashed Blocks}
	\label{blockExample}
	\end{center}
\end{figure}

%%%%%%%%%%%%%%%%%%%%%%%%%%%%%%%%%%%
\vspace{5pt}
\subsection{Overview of Blockchain}
This section details general blockchain architecture and describes Hyperledger Fabric and its role within this work.

\subsubsection{Description and Variations}

Blockchain technology was first introduced by Nakamoto in \cite{bitcoin}. A blockchain is essentially a chain of hashed 'blocks' where each block contains a time-stamp, the previous block's hash, and a collection of transactions. This idea is illustrated in Fig. \ref{blockExample}. Transactions represent many things, but in the Bitcoin architecture, it is a transfer of currency from one digital location to another. In other blockchain implementations these transactions can be an invocation of code stored in the ledger known as 'smart contracts'. A block is generated after a set of transactions have been invoked and are awaiting validation.

There are different types of blockchains and each supports a different level of privacy and security for different use cases. Below is a description of the different types of blockchains as described in \cite{blockchainSurvey}:

%%%%%%%%%%%%
\vspace{5pt}
\noindent\textbf{Public Blockchain:}
In a public blockchain all miners participate in the consensus determination process and the ledger is completely visible to all participants. Public blockchains are permissionless and do not implement access control regarding transaction acceptance. 

%%%%%%%%%%%
\vspace{3pt}
\noindent\textbf{Private Blockchain:}
Private blockchains utilize a centralized architecture where one business or entity controls all of the nodes in the blockchain and writes and validates all transactions. This allows higher efficiency and strict permissions on who can participate in the network. However, all of the flaws that accompany centralization remain.

%%%%%%%%%%%%
\vspace{3pt}
\noindent\textbf{Consortium Blockchain:}
In a consortium blockchain only trusted nodes can participate in the validation of blocks but these trusted nodes are not defined to a single organization or entity. This can provide some of the benefits of the private blockchain such as efficiency and privacy of transactions without compromising the decentralized nature of the public blockchain.

\begin{figure}[t]
    \begin{center}
        \includegraphics[width=\columnwidth]{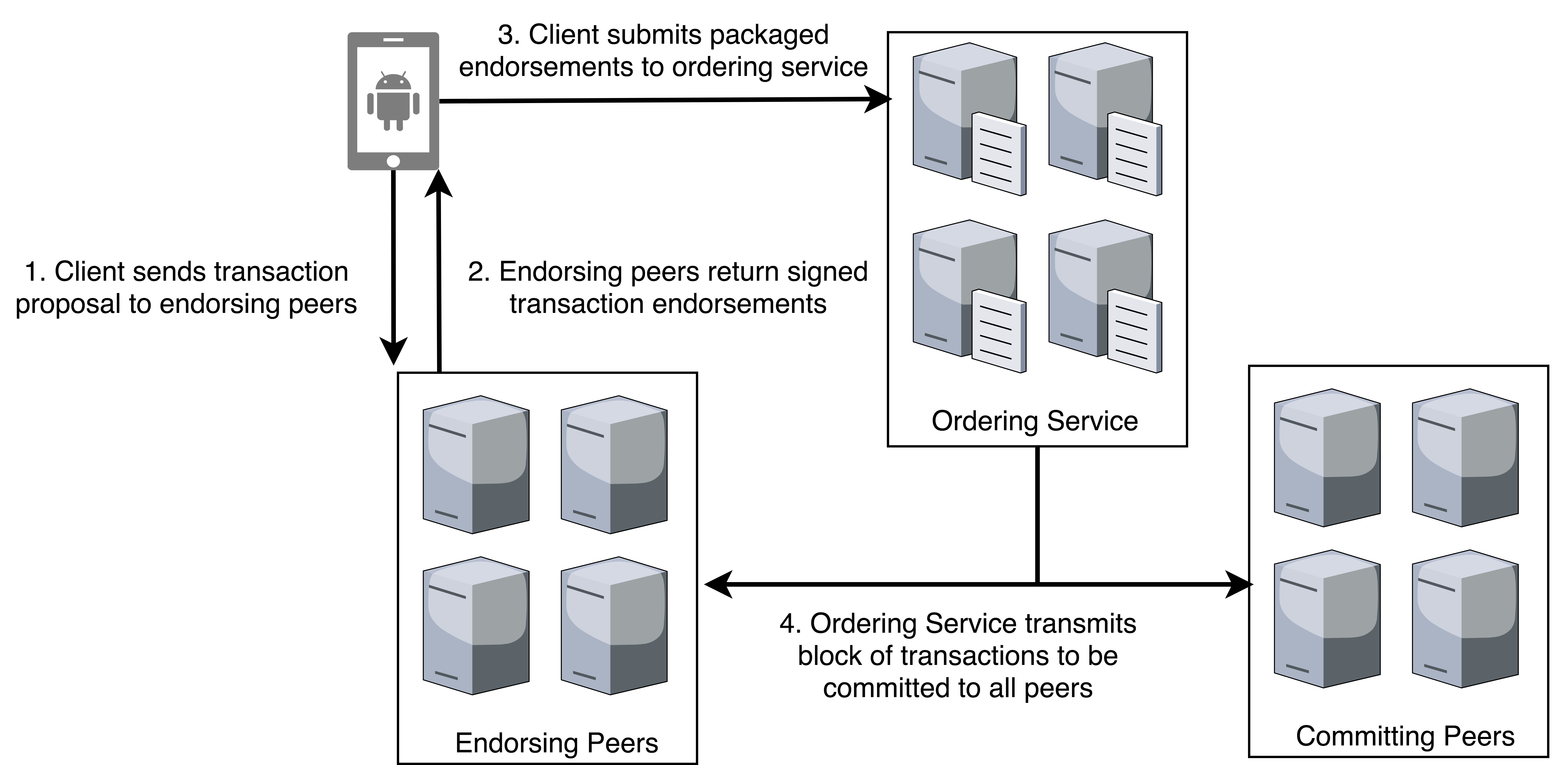}
        \caption{Hyperledger Fabric Transaction Flow}
        \label{hlfTransactionFlow}
    \end{center}
\end{figure}

\subsubsection{Hyperledger Fabric}
A consortium blockchain technology known as Hyperledger Fabric \cite{hlf} is utilized in the implementation and framework proposed in this paper. In Hyperledger Fabric, nodes must be certified before they can participate in the network. However, the nodes are not necessarily owned by one entity. Hyperledger Fabric supports smart contracts (termed ``chaincode" in Hyperledger Fabric) that can be written in any programming language which define all allowable interactions within the network. Each chaincode function has access control functionality such that only certain users / peers can invoke it.
Hyperledger Fabric ``organizations" are used to logically associate peers and users in a cohesive manner to reduce access control overhead. The connections between organizations are created and maintained by a group of ordering nodes termed an ordering service. The ordering service receives certifications from all participating organizations which are used to facilitate permissioned inter-organizational communication via network channels.  

When chaincode is installed on a peer it becomes an ``endorsing peer" with the responsibility of validating proposed transactions that invoke functions of the chaincode. When an endorsing peer validates a proposed transaction it returns an ``endorsement" to the invoking user which contains the endorsing peers cryptographic signature to mitigate falsification. The user must receive a minimum number of endorsements (specified during chaincode deployment) prior to submission to the ordering service. The ordering service packages received transaction proposals into blocks according to a modular algorithm determined at channel creation. Blocks are sent to all peers participating in the channel to be committed to the ledger after one more check for validity. The complete flow from validating a transaction proposal to committing a block is illustrated in Fig.~\ref{hlfTransactionFlow}.

%%%%%%%%%%%%%%%%%%%%%%%%%%%%%%%%%%%
\subsection{Ride-Sharing / Ride-Hailing Background}

In mainstream media when ride-sharing is discussed the term is often used incorrectly. The term ride-hailing describes companies such as Uber and Lyft where a rider requests a specific ride from their current location to a specified destination. The term ride-sharing formally defines situations where a rider accompanies a driver for a portion of a pre-planned trip that was being driven regardless. The implementation described in this paper is directed towards ride-hailing platforms but could be translated to a ride-sharing scenario.

Dynamic ride-sharing describes the problem space of routing for independent rides where routes must be calculated at the time of request rather than beforehand. This can be a challenging topic for optimization due to the lack of internal structure that other forms of ride-sharing such as buses and trains benefit from. There are many variables that must be taken into account such as ride distance, rider wait time, and total number of rides given.

The framework and implementation proposed in this paper do not directly contribute to the optimization of ride-hailing and routing algorithms but rather the underlying ride-hailing architecture. Decentralization of the ride-hailing platform is the main goal of this research but for the implementation to be successful it must provide adequate service to both drivers and riders. Effective routing protocols are essential to providing ease of use to consumers.

%%%%%%%%%%%%%%%%%%%%%%%%%%%%%%%%%%%%%%%%%%%%%%%%%%%%%%%%%
\subsection{Research Challenges and Objectives}

Maintaining a ride-hailing platform independently is costly and with the addition of AVs it becomes prohibitively expensive for most. The alternative architecture proposed in this paper allows individuals to lease their AVs to a community pool where service can be provided to a larger network of users. AV owner's gain the benefit of having control over their organization within the overarching network without having to create an independent application and reach a critical mass of users to achieve profitability.  Incentive is provided to AV owners through increased profit per ride with no percentage to pay to a parent company and through a growing user base shared with competing organizations.  This provides a greater degree of flexibility than can be achieve in a traditional centralized architecture.  This alternative design presents several benefits but also unique challenges such as the lack of trust between users and decentralized infrastructure oversight.

Permissioned blockchain technology provides a solution to these problems by allowing AV owners to securely participate in the network while sharing the burden of maintaining the infrastructure. Hyperledger Fabric provides this functionality and our implementation intends to ease adoption of this technology. In our framework AV owners can join together to create Hyperledger Fabric organizations where they have control over factors such as infrastructure costs and profit distribution. Reliability of the network can be maintained at a large scale once multiple organizations are participating in the network. The infrastructure becomes distributed in this manner and information security can be provided due to the dissemination of ride information being limited to peers participating in a transaction. Currently transportation is mostly an independent consideration where owning a personal vehicle is a necessity outside of urban locations. Adoption of AVs would allow for the normalization of vehicle sharing which could in turn reduce the environmental impact of the automotive industry, provide a financially practical form of transportation to users, and generate profits for AV owners.

Information sharing within the network between actors must be based on necessity for the sake of security. Chaincode permissions provide a mechanism for restricting what information can be queried by entities in the network which is critical to the goal of information security. Riders should not have authorization to access information related to rides that they did not actively participate in. In a co-rider scenario each rider should only have access to information that they were present to observe. For example, co-rider pickup and dropoff locations should only be accessible to a rider if they were present when the information was recorded.

Certifications are used to determine the identity of entities within the network and this determines what resources they are allowed to access such as specific chaincode functions. This necessitates the selection of an appropriate certificate authority implementation that allows for decentralization as well as strong security with regard to authentication and identification. In this work each Hyperledger Fabric organization implements their own certificate authority system to certify its peers, orderers, and users. This ensures that there is no dependency on a centralized certificate authority for the system to work, preserving the distributed nature of the application. The default certificate authority implementation proposed in this paper utilizes the cryptogen tool that is part of Hyperledger Fabric to generate certificates and then manually distributes these certificates. A proper certificate authority configuration is needed to deploy the proposed implementation in a real-world environment. 

%%%%%%%%%%%%%%%%%%%%%
\section{Related Work}
\label{Sec:Related-Work}

%%%%%%%%%%%%%%%%%%%%%%%%%%%%%%%%%%%%%%%%%%%%%%%%%%%%%%%%%%%%%%%%%%%%%%%%%%%%%%%%%%%%%

This section will be used to detail the research that has previously been done in this field or closely related fields. 

%%%%%%%%%%%%%%%%%%%%%%%%%%%%%%%%%%%%%%%%%%
\subsection{Ride Hailing Privacy}
Traditional ride-hailing services such as taxis did not require strong privacy protection as riders could remain relatively anonymous during transactions. With the advent of application based ride-hailing, privacy is a much larger concern. Pham \etal in \cite{pham2017privateride} propose a framework for preserving location privacy of riders and drivers without compromising on functionality. Pham \etal expanded this framework in \cite{pham2017oride} by increasing privacy and addressing the issue of user accountability that can be abused with anonymity. \cite{pham2017oride} allows the service provider of the ride-hailing service to revoke a user's anonymity should they abuse the system. 

Cao \etal in \cite{cao2018practical} address transactional privacy issues by proposing a protocol framework that allows for anonymous ride-hailing and payment for services. This system uses public-key cryptography, an online transportation network, and a third-party payment platform to achieve this. Aïvodji~\etal~in~\cite{aivodji2018sride} propose a privacy-preserving ride-sharing system which protects the privacy of users from the service provider during the matching phase of the ride-sharing system. Implementation of the privacy protecting protocols proposed in these frameworks will be important to the framework proposed in our paper due to the decentralization of the service provider.

%%%%%%%%%%%%%%%%%%%%%%%%%%%%%%%%%%%%%%%%%%
\subsection{Smart Contract Security}

Kosba \etal in \cite{kosba2016hawk} describe ``Hawk" a decentralized smart contract system that does not store sensitive transaction data such as financial information in cleartext in the blockchain. This formal model allows for development of decentralized applications that utilize the blockchain without having to implement encryption within the application. This is very relevant to this paper as one of our primary challenges is securely transferring transaction data; however, Hawk is more focused on public blockchains where all information in the blockchain is generally transparent.

Dubovitskaya \etal in \cite{healthcareHyperledger} propose a blockchain-based system for storing and sharing electronic healthcare records. In the paper the implementation is done using Hyperledger Fabric due to its permissioned nature and access control is implemented through chaincode permissions. The architecture proposed by Dubovitskaya \etal is similar in nature to the architecture proposed in this paper with differences that better suit the healthcare domain. McCorry \etal in \cite{mccorry2017smart} propose additions to previous blockchain-based voting protocols to ensure privacy of voting identities as well as avoiding situations where final voters can calculate results before voting. McCorry \etal also discuss design choices that were made to specifically avoid Ethereum specific smart contract vulnerabilities such as reentrancy attacks and replay attacks. Atzei \etal in \cite{surveyEthereumAttacks} also discuss these attacks among others on the Ethereum environment in their taxonomic aggregation of Ethereum vulnerabilities and poor programming practices.

Delmolino \etal in \cite{smartContractMistakesAndBugs} discuss common smart contract development pitfalls as well as their smart contract security education efforts. This paper discusses some smart contract programming pitfalls that are common to any smart contract development such as logical errors and a lack of data encryption. Ethereum specific mistakes are presented in this paper as well as techniques to avoid / correct them. Luu \etal in \cite{makingContractsSmarter} discuss common avoidable vulnerabilities in Ethereum such as Transaction Ordering Dependence, Timestamp Dependence, Mishandled Exceptions, and the Reentrancy Vulnerability. While some of these vulnerabilities are not directly tied to Hyperledger Fabric it can be useful to learn how mistakes are exploited in other blockchain environments to learn how to better protect a permissioned blockchain.

%%%%%%%%%%%%%%%%%%%%%%%%%%%%%%%%%%%%%%%%%%
\subsection{Blockchain Ride-Sharing}

In \cite{chainedOfThings}, Mehedi Hasan \etal describe a framework for an omnipurpose dependable blockchain to be used as the communication platform for an autonomous vehicle ride-sharing system. 
%Our work is a large extension to this paper.
%
%However, we do not utilize an omnipurpose dependable chain rather we focus on building a specific blockchain implementation for the purpose of a decentralized ride-hailing platform. 
There are several cryptocurrency-based decentralized ride-sharing efforts either currently in development or that have been developed and are in the market as of right now, such as \cite{ridesharingCoinDacsee}, \cite{ridesharingCoinIRide}, \cite{ridesharingCoinBitcab}, \cite{ridesharingCoinRidecoin}, \cite{ridesharingCoinLazooz}, \cite{ridesharingCoinArcadeCity}, \cite{ridesharingCoinChasyr}, and \cite{ridesharingCoinRev}. These projects are similar in design to the project described in this paper with the main difference being that these projects are all public blockchain implementations, mostly Ethereum-based. Public blockchains are not ideal for this work due to the need for private information to be shared between smart contracts. The framework proposed in this paper uses Hyperledger Fabric as its blockchain and the infrastructure is maintained by organizations of drivers rather than by using a cryptocurrency where public miners participate in the validation of transactions. This allows for a finer grain of control over transaction privacy and transaction validation.

Yuan~\etal in~\cite{yuan2016towards} address the key research issues in blockchain-based intelligent transportation systems (ITS) and outline a 7-layer conceptual mode to aid in the development of this field. Yuan \etal also perform a case study by identifying the different architectural components of the most successful blockchain-based ITS known as La'zooz (as described above) and mapping these components to their conceptual model.

The existing research that is most related to the research presented in this paper is the work done by Mehedi~Hasan~\etal in~\cite{chainedOfThings} and the public blockchain ride-hailing implementations. This paper differs from the work of Hasan \etal due to the differing network structure, chaincode protocol, and client application implementation. The structure of the implementation of a decentralized ride-hailing platform proposed by Hasan \etal focuses on the omnipurpose dependable chain used to support the network. Our research could be integrated with an omnipurpose dependable chain but is a deeper dive into the chaincode protocol which facilitates driver / rider communications and the security of these functions. Our protocol is designed such that for a user to transact upon the blockchain they must have a valid certificate and password and returned information will only be relevant to the requesting user.  Blockchain permissions are also exploited to further ensure tampering of data is impossible.  The other current blockchain implementations of a ride-hailing platform are all done utilizing a public blockchain and utilize a cryptocurrency. The usage of a public blockchain leaves room for improvement in the aspect of transaction privacy and authorized network participation.  Ethereum transactions do provide user anonymity to a degree but the information within transactions such as driver / rider locations must also be protected. For example, on a public blockchain any user may view any block that has been committed which could lead to dangerous situations such as a malicious actor tracking the locations of a driver / rider in order to physically harm or steal from them.

%%%%%%%%%%%%%%%%%%%%%
\section{Framework}
\label{Sec:Framework}

\begin{figure}
    \centering
    \begin{subfigure}[]{0.6\columnwidth}
        \includegraphics[width=\columnwidth]{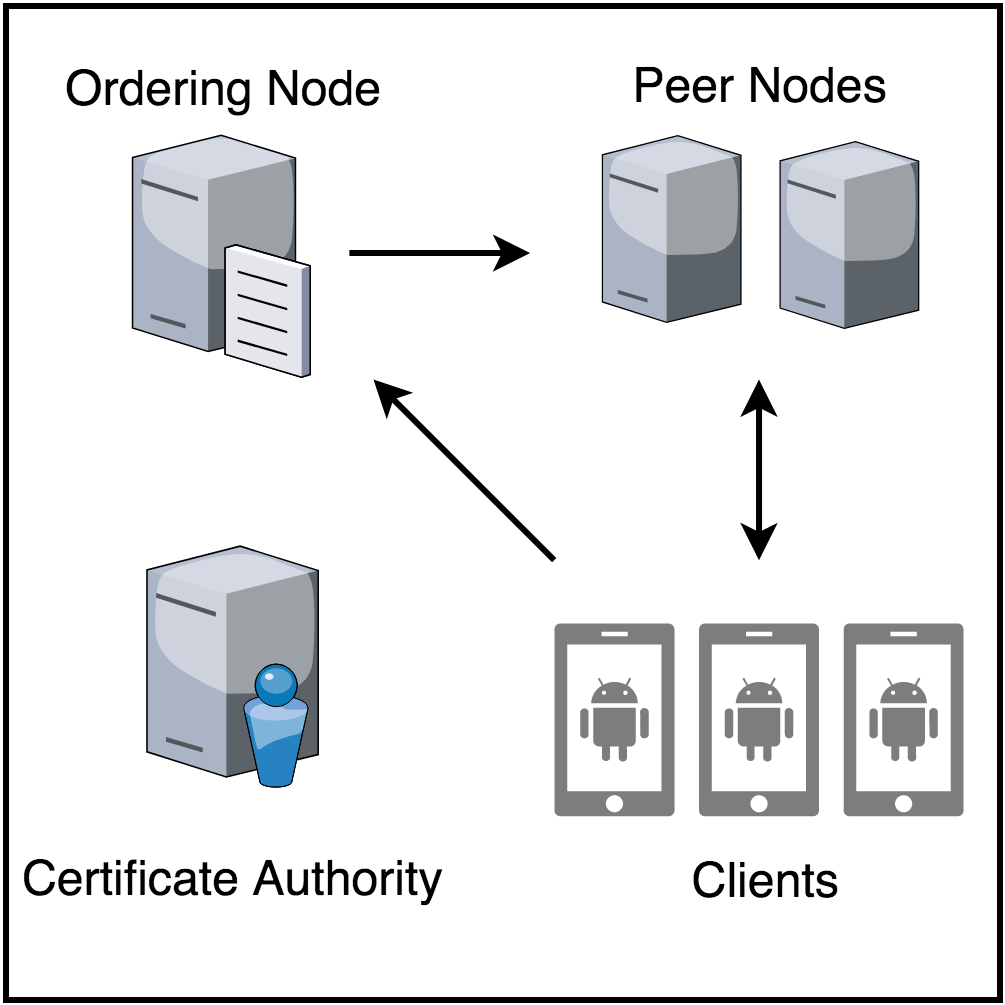}
        \caption{Components of a Hyperledger Fabric Organization}
        \label{organizationStructure}
        \vspace{12pt}
    \end{subfigure}
    ~ %add desired spacing between images, e. g. ~, \quad, \qquad, \hfill etc. 
    \begin{subfigure}[]{0.8\columnwidth}
        \includegraphics[width=\columnwidth]{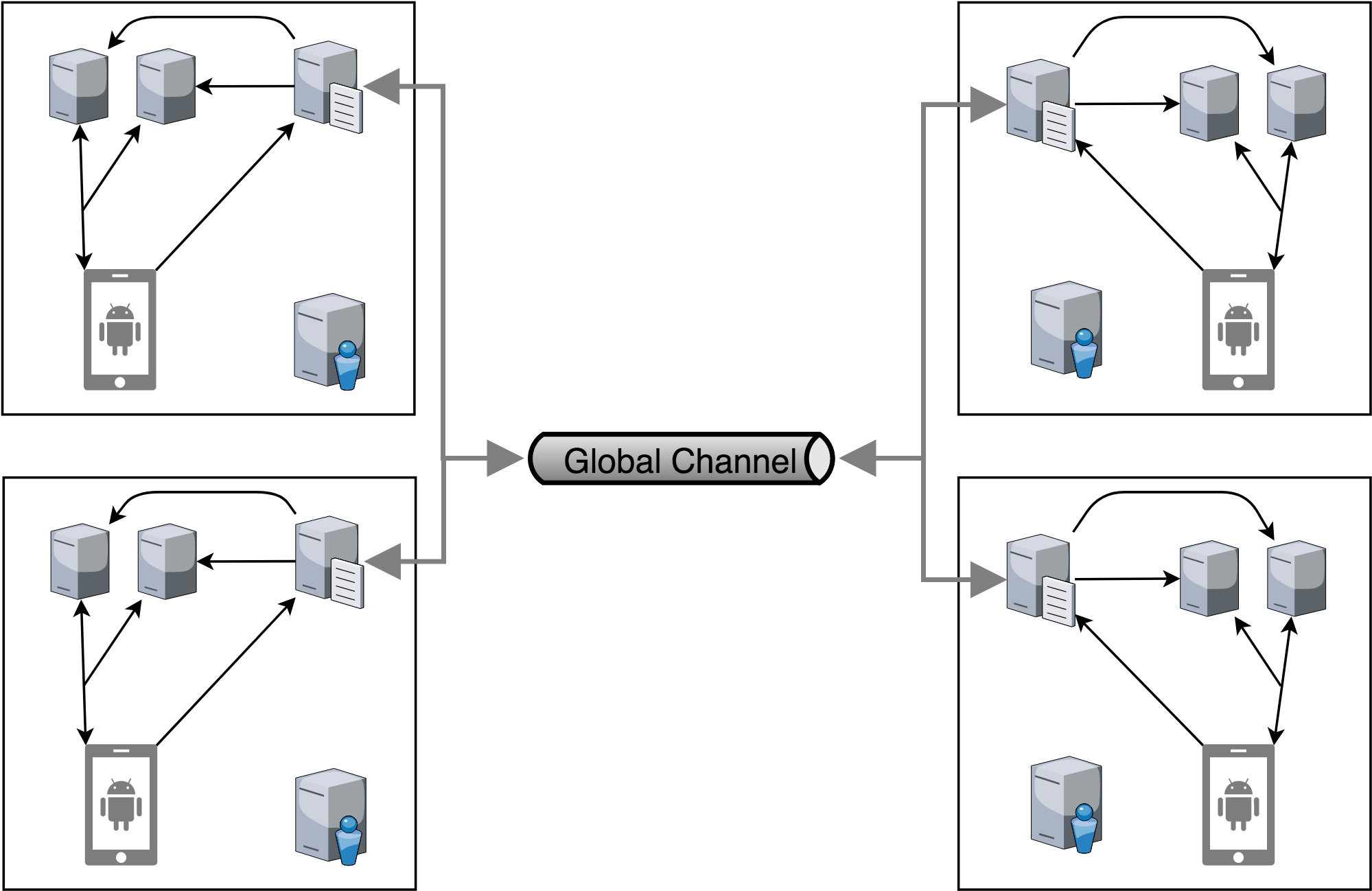}
        \caption{Communication Structure in Multi-Organization Hyperledger Fabric Network}
        \label{networkStructure}
        \vspace{4pt}
    \end{subfigure}
    \caption{Proposed Hyperledger Fabric Implementation}\label{fig:implementationStructure}
\end{figure}

%%%%%%%%%%%%%%%%%%%%%%%%%%%%%%%%%%%%%%%%%%%%%%%%%%%%%%%%%%%%%%%%%%%%%%%%%%%%%%%%%%%%%

The core components of the Hyperledger Fabric architecture that support the ride-hailing framework proposed in this work are: Organizations, Endorsing Peers, Channel Structure, Chaincode Function Structure, Driver / Rider Clients, and Certificate Authorities. The following subsections will describe the proposed blockchain ride-hailing framework and the secure transaction protocol.

%%%%%%%%%%%%%%%%%%%%%%%%%%%%%%%%%%%%%%%%%%%%%%%%%%%%%%%%%%%%%
\subsection{Architecture}

%%%%%%%%%%%%%%%%%%%%%%%%%%%%%%%%%%%
The Hyperledger Fabric organizations used in this framework are created and maintained by groups of drivers / AV owners. These organizations have several core components and this section will detail those components and how they interact to provide ride-hailing functionality while protecting the confidentiality of users. Fig. \ref{organizationStructure} illustrates each of the components required by an organization in the proposed~framework.

%%%%%%%%%%%%%%%%%%%%%%%%%%%%%%%%%%%
\subsubsection{Peer Nodes}
Organizational peer nodes will act as both endorsing peers and committing peers. Endorsing peers are sent transaction proposals from driver and rider client applications in the network and return signed proposal responses. Proposal responses are signed cryptographically to minimize falsification and are marked accepted or rejected based on the transaction validity. After transactions have been ordered and validated, all committing peers in the channel commit the transaction to their local ledger.

Hyperledger Fabric supports multiple channels within a single network. Each channel in this case maintains a completely separate ledger from the other channels in the network and this is used for data segregation. As seen in Fig \ref{networkStructure} this work does not utilize the multiple channel option of Hyperledger Fabric but rather utilizes chaincode permissions and structuring to segment data from unauthorized entities.

%%%%%%%%%%%%%%%%%%%%%%%%%%%%%%%%%%%%%%
\subsubsection{Certificate Authority / Orderer Node}

The root certificate authority of an organization issues certificates to all of the entities within the organization which are the orderer node(s), the peer nodes, and all of the end-user client applications (drivers and riders). These certificates are stored locally on each individual entity within the local MSP. Upon creation of the ordering service all ordering nodes share local MSP details between one another. This allows all of the organizations in the network access to the certificates that can be used to validate the identity of entities in the network. This is done so that endorsing peers can validate that the user invoking a chaincode function is both certified through an organization and is authorized to access the specified chaincode. Any certificate authority implementation can be used for this purpose but a client / server architecture is recommended by Hyperledger~Fabric.

%%%%%%%%%%%%%%%%%%%%%%%%%%%%%%%%%%%%%%%%%%%%%%%%%%%%%%%%%%%%%
\subsection{Chaincode Protocol}

Careful design of the chaincode functions that are installed on endorsing peers is critical to the goal of securely storing rider information in a manner where it is accessible only to the rider it pertains. The chaincode can be written in any language but it is important to note that the proposed implementation was written in golang and is dependent on certain functions included in the Hyperledger Fabric core go library \cite{fabric}. Future projects that utilize this framework with chaincode written in another programming language must find appropriate substitutions for this dependent functionality. 

Distinction between individual riders and drivers is done within the chaincode by creating unique UserIDs and then utilizing these IDs within the chaincode functions. A UserID is created by concatenating the rider or driver's local MSP ID with the organization's global MSP ID. Each local MSP ID is generated when the entity is registered with the organization MSP and is guaranteed to be unique within the organization. The global MSP ID and the local MSP ID are both tied to the certificate that the user implicitly passes to the chaincode. It is impossible to impersonate another system user without having access to their certificate and password. This ensures security of user information assuming the chaincode framework is properly used and login functionality is implemented in the client application.

\begin{figure}[t]
	\begin{center}
	\includegraphics[width=0.8\columnwidth]{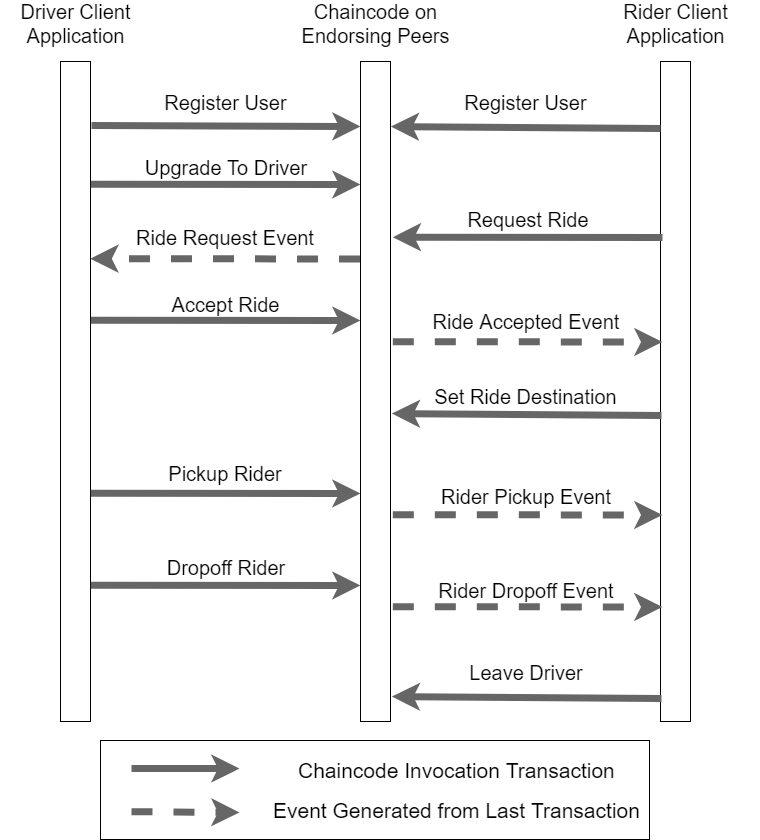}
	\caption{Ride-Hailing Protocol Framework}
	\label{hlfFrameworkProtocol}
	\end{center}
\end{figure}

%%%%%%%%%%%%%%%%%%%%%%%%%%%%%%%%%%%%%%%%

The following functions provide the core functionality of the proposed framework and will be called by the client application by both drivers and riders to facilitate the ride-hailing process:

\vspace{5pt}

\noindent\textbf{registerUser / unregisterUser:} Creates a new user object in the ledger using the unique UserID as the key and the function parameters for values. The values that are attached to a unique user are the hash and salt of the rider's password and an array of ride structs which will be filled as the user provides or requests rides. Unregister user needs to delete this key from the ledger. In our implementation we provide another chaincode function for users to add their name and provide extra information to become a driver. This functionality could be included in this chaincode function if desired.

\noindent\textbf{requestRide:} Creates a temporary ledger value for the ride that is being requested where data will be stored until the end of the ride. Each participant will retrieve the information relevant to them to be stored permanently in the ledger at the end of the ride. This function also needs to create an event that will be received by listening drivers (implemented in the client application).

\vspace{3pt}

\noindent\textbf{acceptRide:} Updates the temporary ride object created by the previous function to mark that the ride has been accepted and to designate the accepting driver. This function also creates an event which alerts the requesting rider that a driver is en route.

\vspace{3pt}

\noindent\textbf{setRideDestination:} Updates the temporal ride object to include the ride destination coordinates when the signal from acceptRide is received. This is done after driver acceptance to prevent discrimination based on dropoff location.

\vspace{3pt}

\noindent\textbf{pickupRider:} Called when the driver reaches the rider's location. This function performs checks to ensure the ride is still ongoing and the driver is at the correct location and then triggers an event to alert the rider that the driver has arrived.
\vspace{3pt}

\noindent\textbf{dropoffRider:} Called when the driver reaches the final destination. Pulls necessary information from the temporal ride object to create a permanent ride object specific to the driver and creates an event to alert the rider that the ride is ending.
\vspace{3pt}

\noindent\textbf{leaveDriver:} Triggered by the event created by dropoffRider. Creates a permanent ride object for the rider and transfers necessary information from the temporal ride object which is then deleted because all information has been collected.
\vspace{3pt}

\noindent\textbf{setCo-riderInformation:} This function is called whenever a co-rider joins or leaves the ride where the corresponding rider is present. The co-rider information passed to this function is added to the temporal ride object to be stored permanently later in the ride.
\vspace{3pt}

\noindent\textbf{getUserInfo:} Retrieves the values stored for the calling rider or driver that are set during registration. This functionality is used to retrieve the user's password hash and salt as well as the list of RideIDs associated with this user.
\vspace{5pt}

This core functionality provided by the chaincode is invoked by the client application to provide the base ride-hailing service. The client application implementation could be changed to allow for more functionality but the above chaincode functions provide the core framework. The specific implementation used for this work will be described in the section below. An illustration of the chaincode protocol is depicted in Fig.~\ref{hlfFrameworkProtocol}.

%%%%%%%%%%%%%%%%%%%%%
\section{Implementation}
\label{Sec:Implementation}

%%%%%%%%%%%%%%%%%%%%%%%%%%%%%%%%%%%%%%%%%%%%%%%%%%%%%%%%%%%%%%%%%%%%%%%%%%%%%%%%%%%%%

We implement the framework described above using two organizations both having two peers (both endorsing and committing), an orderer, a certificate authority, and then multiple user accounts which are utilized by the client application to register drivers and riders. It was decided that organizations should maintain a minimum of two peer nodes so that a singular organization can operate independently while still providing fault tolerance. There is not a formal certificate authority node within the organizations in this implementation but rather cryptogen (the certificate tool provided by Hyperledger Fabric) was integrated into the build process so certificates were generated and distributed manually during network initialization. In this implementation new users have to be generated manually and can be done after the network is started; however, the functionality is not included within the client application. The client application is built using golang but could be built using any language (particularly languages with already available Hyperledger Fabric client SDKs). We utilize the fabric-sdk-go library \cite{fabricSdkGo} to communicate with the endorsing peers of our network. 

The peers of this network are built using docker~compose~\cite{dockerCompose} and run within separate docker \cite{docker} containers in a single docker network. The client application currently runs on the host machine and accesses the Hyperledger Fabric network through ports that have been exposed to the host machine via docker. In a production deployment the peer, orderer, and CA nodes would be servers accessible via WAN. These servers can all be located within the same machine and could still run in docker (which is ideal for fault tolerance) but they need to be accessible via defined sockets separate from one another. The client application GUI was built using a golang wrapper for the Qt cross-platform application development library~\cite{qt}~\cite{goQt}. This was done so the application could be developed for desktop and easily ported to android devices. 

\begin{figure}
    \centering
    \begin{subfigure}[]{0.4\columnwidth}
        \includegraphics[width=\columnwidth]{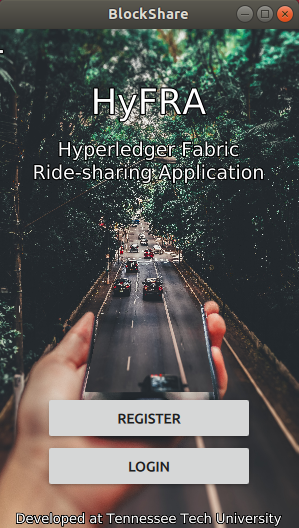}
        \caption{Main Menu Functions}
        \label{mainMenuFigure}
    \end{subfigure}~
    \begin{subfigure}[]{0.4\columnwidth}
        \includegraphics[width=\columnwidth]{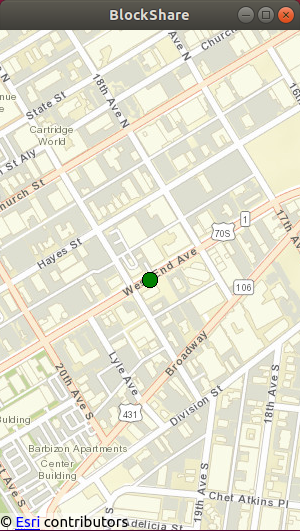}
        \caption{Driver Map Window}
        \label{driverMap}
    \end{subfigure}
    \caption{Application Screenshots}\label{fig:screenshots}
\end{figure}

When the client application is started the user is presented with a main menu with basic login functionality as shown in Fig. \ref{mainMenuFigure}. When the user navigates to the register button they are prompted for the information that is needed to create the ledger object as described in the framework section. Once the user fills the input fields and continues, the password provided is salted and hashed and the registerUser chaincode function is called. 

The login functionality of the client application prompts the user for their username, organization, and password. The rider is also given an option to authenticate as a rider or a driver. An option to upgrade to a driver is given after login. Upon receiving the user's credentials the application calls the getUserInfo chaincode function and retrieves the stored hash and salt values for verification. If the user is logging in as a driver they are presented with a startDriving function and a logout function. The startDriving function is the core of the driver's client application experience and starts the ride-hailing process. When the startDriving function is activated the user is prompted to provide the address of their current location. Once this is provided geolocation is utilized to convert this to a latitude and longitude. The geolocation service as well as the mapping service described below is provided through a plugin for Qt which interacts with the open-source project named Open Street Map \cite{openStreetMap}. Once a latitude and longitude for the driver is obtained the driver's application updates with a map of their current location and the driver starts listening for events on the requestRide chaincode. The starting driver map can be seen in Fig. \ref{driverMap}.

When a rider authenticates they are presented with the option to request a ride, update their profile, upgrade to driver status, or logout. If the user wishes to become a driver they must provide additional information such as their vehicle make, model, and year. If the user opts to request a ride he/she is prompted to enter their starting location and ending location addresses and the same geolocation process as before takes place. Once the starting latitude and longitude coordinates are calculated the requestRide chaincode function is called and the user listens for events on the acceptRide chaincode function. When a driver is found the rider sends their destination coordinates to the setRideDestination chaincode function where the temporal ride object will be updated. Upon arrival, the driver's client application updates the temporal ride object. As the driver is moving from the rider's pickup location to the newly added dropoff location the driver listens for new ride requests. If a new ride request is received the driver accepts the ride in the same manner as before and the original rider receives an event which invokes the addCo-riderInformation chaincode function. When a ride ends the dropoffRider chaincode function is called which finalizes ride information and generates an event with the RideID. All co-riders use the RideID to update their co-rider dropoff location in their temporal ride object.

%%%%%%%%%%%%%%%%%%%%%
\section{Case Study}
\label{Sec:Case Study}

\begin{figure}[t]
	\begin{center}
	\includegraphics[width=\columnwidth]{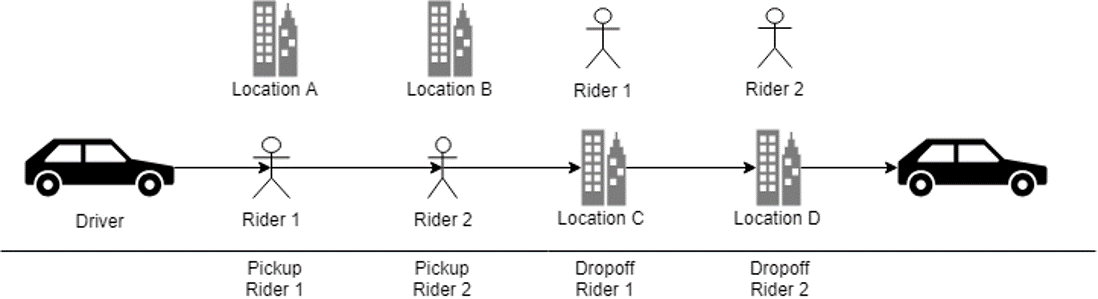}
	\caption{Simple Co-Rider Situation}
	\label{simpleCoriderFigure}
	\end{center}
\end{figure}

%%%%%%%%%%%%%%%%%%%%%%%%%%%%%%%%%%%%%%%%%%%%%%%%%%%%%%%%%%%%%%%%%%%%%%%%%%%%%%%%%%%%%%%%%%%%%%%%%%%%%%%%%%%%%%%%%%
This section will define a typical use case and show the privacy protections in place regarding data stored on the ledger.

%%%%%%%%%%%%%%%%%%%%%%%%%%%%%%%%%%%%%%%%%%%%%%%%%%%%%%%%%
\subsection{Simple Co-rider Scenario}
A simple ride-hailing situation is shown in Fig. \ref{simpleCoriderFigure} where two riders (\textit{R$_{1}$} and \textit{R$_{2}$}) request a ride from driver \textit{D$_{1}$}. In this scenario \textit{R$_{1}$} is present for \textit{R$_{2}$}'s pickup and \textit{R$_{2}$} is present for \textit{R$_{1}$}'s dropoff. This information should be reflected during the execution of transactions within the Hyperledger Fabric network. This section will be detailing these transactions in the proposed implementation by using real locations and showing the contents of the ledger that are accessible by each respective entity. The resulting ledger query from \textit{R$_{1}$} should show the pickup location for \textit{R$_{2}$} but not the dropoff location and vice versa for the query by \textit{R$_{2}$}. The driver's ride object in the ledger should contain the pickup and dropoff locations for both riders as they were present for all events.

\begin{figure}[t]
	\begin{center}
	\includegraphics[scale=0.8, keepaspectratio=true]{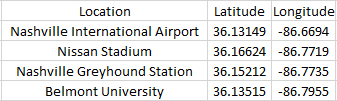}
	\caption{Case Study Location Coordinates}
	\label{locations}
	\end{center}
\end{figure}

The locations used for this scenario are all public locations in Nashville, TN and represent a typical use case for the ride-hailing platform's operation. The driver starts listening for new ride events while located in downtown Nashville. \textit{R$_{1}$}'s starting location is Nashville International Airport and his destination is Nissan Stadium. \textit{R$_{2}$}'s starting location is the Nashville Greyhound station and their destination is Belmont University. The latitudes and longitudes for these locations are show in Fig. \ref{locations}. The locations are stored in latitude / longitude pairs in the ledger so this table will be important as a reference.

When the driver elects to start listening for ride requests they are shown a screen similar to Fig. \ref{driverMap} with a marker on their current location. In a final build of the application this screen would update with the location pulled from the mobile device GPS. For development purposes, the coordinates are entered when driving is initiated and remain static until a ride is received. At this point in time the driver has authenticated with the Hyperledger Fabric network and registered for chaincode events on the requestRide chaincode function.

\begin{figure}[t]
	\centering
	\captionsetup{justification=centering}
	\includegraphics[width=0.8\columnwidth]{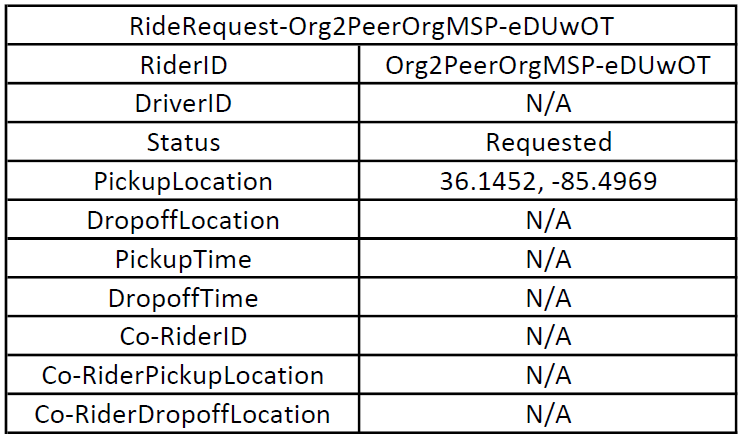}
	\caption{Temporal Ride Request Created in Ledger When Rider-1 Requests a Ride}
	\label{startingRideRequest}
	\centering
	\captionsetup{justification=centering}
\end{figure}

The next step for this scenario is \textit{R$_{1}$} opts to request a ride in their application. The rider enters a starting address and ending address which are converted to latitude and longitude using geolocation. The client application takes this latitude and longitude and calls the requestRide chaincode where the temporal ride request key in the ledger is created which can be seen in Fig. \ref{startingRideRequest}. As seen in Fig. \ref{startingRideRequest} the ID of the key in the ledger is built using values specific to the rider. Org2PeerOrgMSP is the unique MSP name and eDUwOT is the rider's unique ID within the MSP. For this specific ride request to be generated this user must call the chaincode because these two values are generated using the certificate that is passed by the calling user. This is a key part of how the security of this architecture functions. The values that are missing are filled in as the ride progresses and this temporal ride object is used to create the permanent entries in the ledger before it is deleted at the end of the ride. The rideRequest chaincode being called also sends an event to drivers that are listening, providing the rideRequest ID. When a driver accepts the ride the status field of the temporal ride request object is updated as ``accepted". This allows for other drivers to be notified that the ride has already been accepted.

In this scenario when the driver arrives at \textit{R$_{1}$}'s pickup location and begins moving towards the destination another ride is requested by \textit{R$_{2}$}. The driver is then prompted on his application to either accept or deny this ride. In this scenario they accept the ride which updates the current destination. When \textit{R$_{2}$} requested the ride another temporal ride request key was instantiated (there can be exactly one active for each unique rider) with the starting information similar to Fig. \ref{startingRideRequest} but with the values being related to \textit{R$_{2}$}.

When the driver reaches \textit{R$_{2}$} for pickup the location must be stored for \textit{R$_{1}$} so that it can referred to later in the permanent copy of this ride's key. \textit{R$_{1}$} was present for \textit{R$_{2}$}'s pickup so this information should be accessible to \textit{R$_{1}$}. The driver stores this information for \textit{R$_{1}$} because he has access to all current ride information. This is done by iterating through the list of rides currently in progress and recording the co-rider pickup location in the temporal rideRequest for each rider. This same type of iteration is done once \textit{R$_{1}$}'s dropoff location is reached except for each rider that is present during a dropoff the Co-rider Dropoff Location field is filled.

\begin{figure}[t]
	\centering
	\captionsetup{justification=centering}
	\includegraphics[width=0.8\columnwidth]{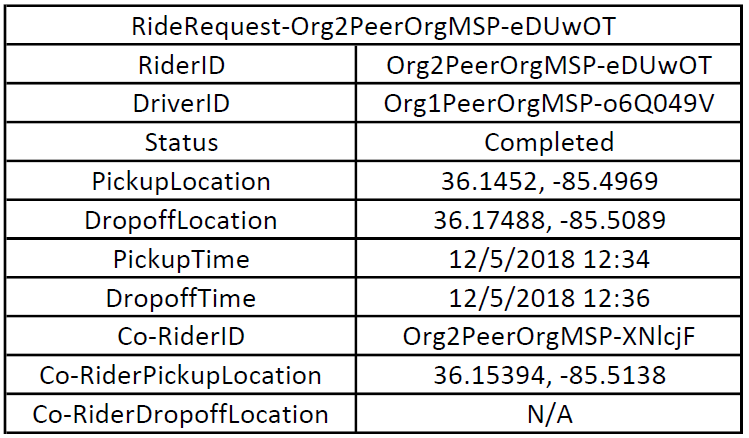}
	\caption{Completed Temporal Ride Request for Rider-1 Before Deletion}
	\label{endingRideRequest}
\end{figure}

Fig. \ref{endingRideRequest} shows \textit{R$_{1}$}'s temporal rideRequest object prior to relocation to permanent storage. The Co-rider Pickup Location field has been filled because \textit{R$_{1}$} was present for this portion of the ride and therefore the information is relevant to \textit{R$_{1}$}. \textit{R$_{2}$}'s final temporal rideRequest was very similar with the exception that \textit{R$_{1}$}'s dropoff location was recorded. The driver handles the archiving of co-rider pickup and dropoff locations because he has access to all of this information during the ride. After the ride is over the riders can only access information that is directly relevant to them. This architecture can be carried over to other applications as well or could be expanded within this application to store additional sensitive data.

%%%%%%%%%%%%%%%%%%%%%
%\input{SecEvaluation.tex}

%%%%%%%%%%%%%%%%%%%%%
\section{Evaluation}
\label{Sec:Evaluation}

This section details the testing that was performed against the completed implementation to review its security and load resiliency.

%%%%%%%%%%%%%%%%%%%%%%%%%%%%%%%%%%%%%%%%%%%%%%%%%%%%%%%%%
\subsection{Chaincode Analysis}
There are several tools available for the static analysis of smart contracts written for public blockchains such as Ethereum but currently only one static analysis tool for Hyperledger Fabric chaincode written in golang. This tool is developed by ChainSecurity and it is a proprietary tool with the free version only usable for non-commercial use \cite{chainsecurity}. We utilized the ChainSecurity tool to analyze our chaincode for design patterns that result in non-determinism or states that could be exploited by malicious actors in the system. The chaincode deployed in our implementation did not have any flaws according to the analysis done by this tool.

Specifically, the static analysis tool checks for concurrency, unchecked exceptions, ledger operations that depend on the global state, field declarations, blacklisted import statements, reading from the ledger after a write operation, and unsanitized input to the chaincode. Several of the issues that are scanned for are related to non-determinism such as the concurrency check, field declarations, and global state checks. The global state and field declarations being used in ledger operations could be exploited because these global variables are only specific to the peer the chaincode is installed on. If this peer crashes or becomes out of sync with the other peers of the network it will never be able to submit another transaction due to its improper readset. Concurrency is highly discouraged within blockchain applications due to inherent issues such as race conditions that can be mitigated within singular applications but are more volatile in a distributed scenario. Exceptions must be checked within the chaincode because if failure is not predictable it could be used by an attacker to access unauthorized ledger components. The static analysis of the chaincode assures that the implementation will not be susceptible to these types of attacks.

\subsection{Performance Evaluation}
\label{Sec:LoadEvaluation}

\begin{figure}[t]
	\begin{center}
	\includegraphics[width=\columnwidth]{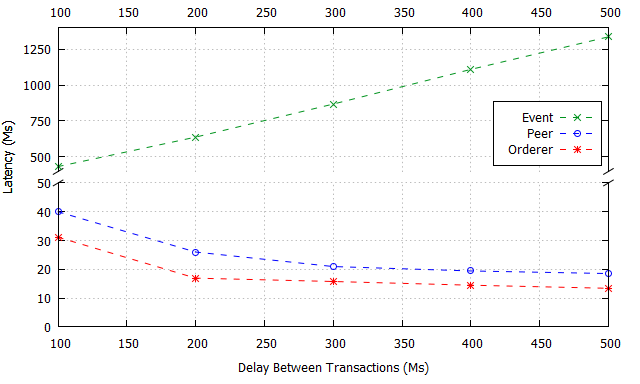}
	\caption{Constant Rate Traffic Load Test}
	\label{constantTrafficTest}
	\end{center}
\end{figure}

\begin{figure}[t]
	\begin{center}
	\includegraphics[width=\columnwidth]{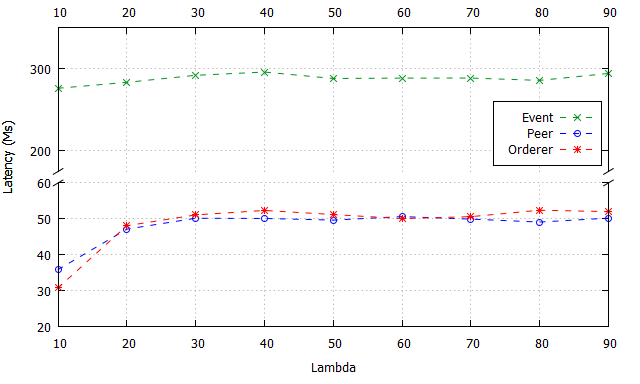}
	\caption{Poisson-Based Traffic Load Test}
	\label{poissonTrafficTest}
	\end{center}
\end{figure}

We utilized a tool provided by Hyperledger Fabric within the fabric-test repository \cite{fabricTest} called the Performance Traffic Engine (PTE) to test our implementation under load. PTE is designed to enable testing of live Hyperledger Fabric networks with various chaincode, orderers, and peers. All test cases processed a total of 6,000 transactions for a total of 1,000 rides with varying transaction rates. Transactions were sent at varying rates according to the test distribution but were handled by four processors with two processors dedicated to each organization. Each processor was responsible for sending 1,500 total transactions per test. Each data point in Fig. \ref{constantTrafficTest} and Fig. \ref{poissonTrafficTest} reflects a single test case. All transactions and events sent were received successfully across all tests. The ordering service used in our implementation has a batch timeout of 2 seconds and a max message count of 10. Whenever one of these limits is reached a block is created. 

There are three values being recorded across all tests: peer, orderer, and event latency. Peer latency measures the amount of time between the moment a transaction proposal is submitted for endorsement and the moment the endorsements are returned to the client. The orderer latency measures the amount of time it takes to receive a transaction acknowledgement from the ordering service after the client submits its endorsements. Event latency measures the time between event registration before submission to the ordering service and a successful block commit. Each value is averaged over 1000 transactions and these averages can be seen in the graphs below.

It is worth noting that in order to test this application at a true scale with thousands of nodes participating in the blockchain would require thousands of machines due to the limiting factor being CPU power.  These tests cannot be virtualized with accuracy because multiple virtual nodes running on a single machine results in a superficial decrease in network performance.  These performance metrics are intended to indicate the increase in load in different scenarios. Due to the design of the network load would not significantly increase past a certain point because the number of nodes validating a single transaction can be reduced to further distribute the workload.

%%%%%%%%%%%%%%%%%%%%%%
\subsubsection{Constant Rate Network Traffic}
The first set of tests sent transactions to the network at varying constant rates with a 30\% deviation. Fig. \ref{constantTrafficTest} shows the results of sending traffic to the network with the minimum delay being 100ms between transactions and with a maximum delay of 500ms. As the delay between transactions increased the event latency also grew at a constant rate. The event latency is measured end-to-end between transactions so the increase is logical due to the artificial delay between transactions. Orderer and peer latency showed an overall decrease as the delay between transactions increased due to the decreasing load being placed on the network.

A second constant rate load test was done with delays ranging between 10 and 90 ms. The event, peer, and orderer latency all remained essentially constant throughout the test due to transactions being sent from one machine using 4 processors. The delay between transactions was not large enough for previous transactions to be received by the testing engine. As the tests approached a delay of 90 ms the peer and orderer delay can be seen to slightly decrease as a more reasonable testing speed is approached.

\subsubsection{Poisson Distribution-Based Network Traffic}
In the interest of testing the network under different types of loads the next test sent transactions according to the Poisson distribution with varying lambda values. The Poisson distribution was chosen because it relates to events that occur independently and is often used to model event functions where the average number of events is known. In our scenario we tested scenarios with varying transactions per second as the average event variable lambda.

Fig. \ref{poissonTrafficTest} shows the results of testing the network with a lower range of lambda values between 10ms and 90ms. Event latency remained mostly constant as the lambda value increased due to the network load increasing and the testing delay decreasing which had the effect of balancing one another. Event and peer latency increased as transaction rate increased due to the increasing load on the network which was the expected result. After reaching a lambda of 30 the peer and event latency leveled off and remained constant. This can be again attributed to the testing engine reaching a point where the system resources were not enough to send transactions at a rate higher than 30 per second. A higher lambda test was conducted as well with lambda ranging between 100 and 1000 but all of the measured latencies remained constant again due to the limitations of system resources. 

\begin{figure}[t]
	\begin{center}
	\includegraphics[width=0.8\columnwidth]{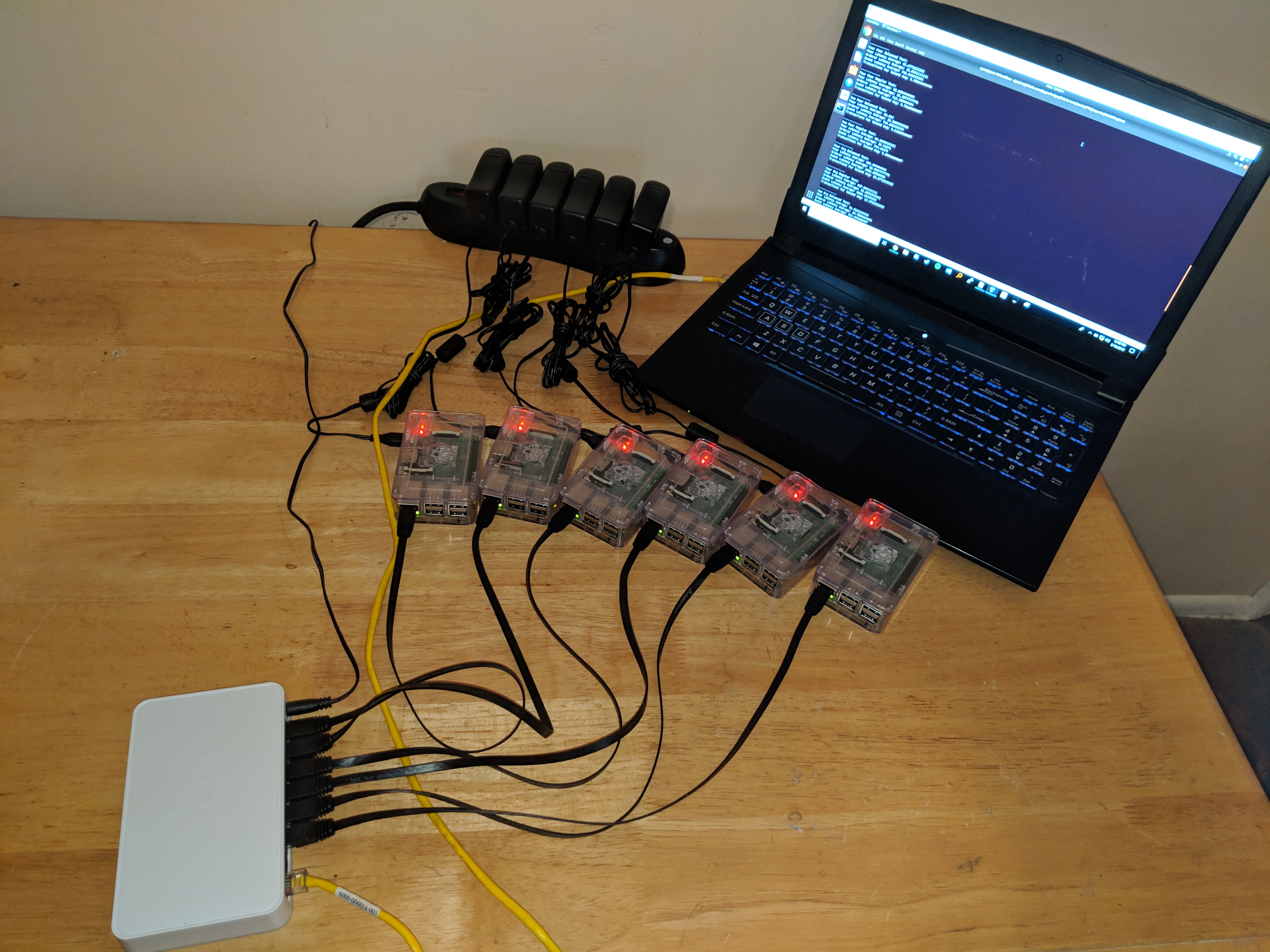}
	\caption{Raspberry Pi Test Bed}
	\label{testbed}
	\end{center}
\end{figure}

\subsubsection{Organization Restructuring Tests}
The final set of tests measured the effect of having different configurations of peer and orderer nodes as well as the effect of having more organizations participate in a network. These tests required more computers so that the results were not limited by the processing power of a single machine. To complete these tests a test bed was built using Raspberry Pi computers \cite{upton2014raspberry} as the host machines for the organizations in the testing. A picture of the test bed is shown in Fig. \ref{testbed}. Each Raspberry Pi was running a Ubuntu 18.04 image that was designed for ARM architectures. In each of the tests in this section one of the Raspberry Pi's acted as a dedicated Kafka / Zookeeper node cluster which the ordering nodes of the network communicated with. The Raspberry Pi machines were connected via a network switch to the testing computer where transactions originated from. Transactions were sent in bursts with a delay of 200ms between them. Each burst of transactions sent two transactions per organization to the respective Raspberry Pi machine. 

\begin{figure}[t]
	\begin{center}
	\includegraphics[width=\columnwidth]{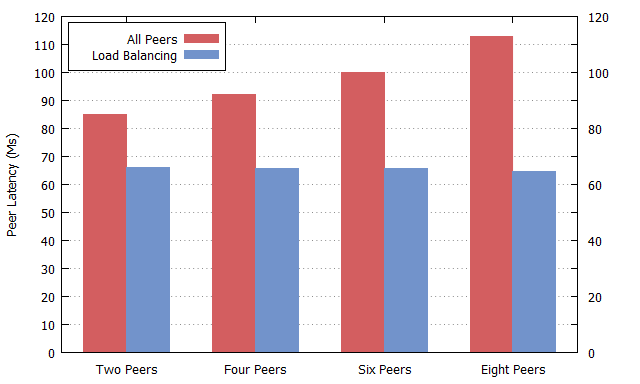}
	\caption{Peer Latency with Differing Numbers of Peers}
	\label{peerReconfig}
	\end{center}
\end{figure}

The first test measured the effect of joining additional peers to a single organization while the amount of traffic being sent remained constant. Neither the speed nor amount of traffic sent to the organization was adjusted between tests and the results can be seen in Fig. \ref{peerReconfig}. Two separate tests were conducted, one required all peers to endorse each transaction and the other balanced endorsements between peer nodes. In the first scenario it can be seen that the peer latency was increasing over time. As each pair of peer nodes was joined to the network there was around a 5-15 Ms increase in peer latency. This is due to the communication overhead that is required by the client to organize the endorsement of their transactions through each peer. The second test utilized peer load balancing where only one endorsement from a peer node was required for a transaction to be considered valid. In this scenario the peer latency can be seen slightly reducing over time because there are more peers than there are transactions being received and therefore the workload on a single peer is never increasing. A very similar test was also performed with orderers being the subject but the traffic level that was used was not enough that the ordering service saw any sort of increase in latency. It is also worth noting that PTE has functionality for orderer node load balancing but this did not have an effect on the latency of the network when included in the tests. All orderer nodes should be participating in the ordering of transactions into blocks so this portion of the test was omitted.

\begin{figure}
    \centering
    \begin{subfigure}[]{\columnwidth}
        \includegraphics[width=\columnwidth]{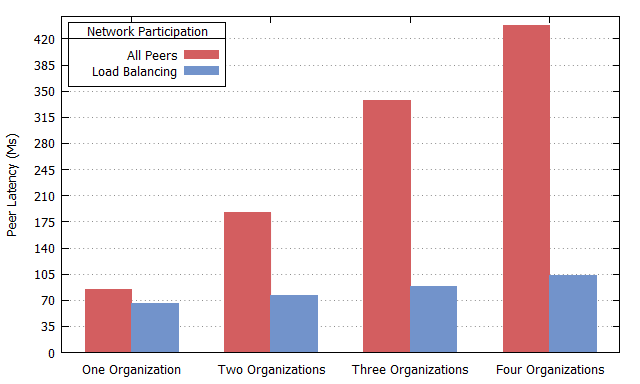}
        \caption{Peer Latency}
        \label{orgReconfigPeerLatency}
        \vspace{4pt}
    \end{subfigure}
    ~ %add desired spacing between images, e. g. ~, \quad, \qquad, \hfill etc. 
    \begin{subfigure}[]{\columnwidth}
        \includegraphics[width=\columnwidth]{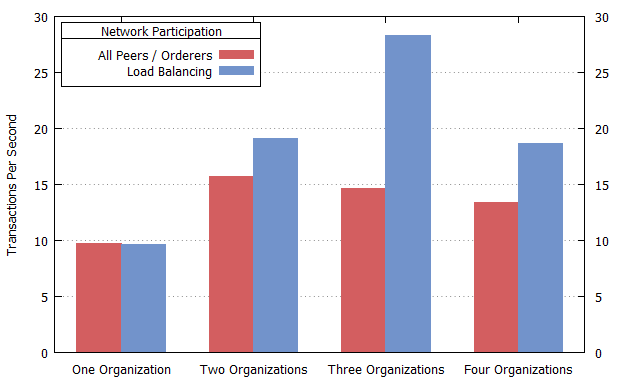}
        \caption{Transactions Per Second}
        \label{orgReconfigTPS}
        \vspace{6pt}
    \end{subfigure}
    \caption{Network Statistics of Organization Reconfiguration Tests}\label{fig:animals}
\end{figure}

The final test increased the number of organizations as well as the amount of traffic in the network at each step to measure the effect on performance. The simulation was again split into two overarching tests where one required all peer nodes to participate in the validation of each transaction and one where the load was distributed evenly between peer nodes. Each organization was hosted on a single Raspberry Pi running two peer nodes and a single orderer node. The peer latency measurements can be seen in Fig. \ref{orgReconfigPeerLatency}. The peer latency and the amount of traffic being sent had a positive correlation due to the work load of each peer increasing at each step in the full peer participation test. During the load balancing simulation it can be seen that the peer latency is also increasing but only by 10-15 Ms at each step. The transactions were being sent in parallel with two transactions being sent per organization every 200 Ms. In the load balancing simulation each thread sending transactions was communicating with a dedicated peer node which explains the much smaller increase in latency.

Fig. \ref{orgReconfigTPS} illustrates the transactions per second (TPS) that were being committed to the ledger. It can be seen again that the load balancing simulation outperforms the full participation simulation. In both tests there is an increase in TPS before it begins to fall. The increase can be attributed to the network not being fully utilized until the network traffic rate is increased to the optimal level. The decrease in TPS comes after the network reaches a critical mass of transactions after which performance drops off. The point was reached in both tests where the peer nodes could no longer process transactions before the next set of transactions were received which formed a bottleneck during the endorsement stage. These transactions are being endorsed on the same machines that must commit the blocks after the ordering stage. Work is constantly being done on the peer node which slows down the performance of the committing stage where additional verification is done. The only way to subvert this bottleneck would be to segregate the endorsing peers from the committing peers which would easily be possible with this architecture; however, it would increase hardware costs for the organization operators.

%%%%%%%%%%%%%%%%%%%%%
\section{Conclusion and Future Work}
\label{Sec:Conclusion}

This paper serves as a framework for building a decentralized ride-hailing application that serves as the intermediary platform connecting drivers and riders. The chaincode protocol described in this paper provides security of transactions through design and could be extended to many applications. Information returned by from the ledger will only relate to the calling user due to the implicate certificate that is passed and further used in the pre-designed transactions proposed in this work.  The implementation utilizes permissioned nature and built-in organization structure of Hyperledger Fabric to detail an optimal build for organizations of independent drivers. If the principles are followed for the chaincode development then the client application that interacts with the chaincode cannot risk the safety of the users in the system. Ideally each organization in this system will have the ability to create their own client application and still be able to interact with the Hyperledger Fabric network and share the load of the riders.  This flexibility would allow driver organizations to truly be in control of their application while also allowing them to grow their user base together by providing a distributed web of service.

The assumption under our architecture is that each organization employs a certificate authority to manage identification of users in the organization (including both drivers and riders). This allows for certification of users to remain implemented without relying on a centralized certificate authority. Certificate authority public key infrastructure is recommended by Hyperledger Fabric but future development could utilize a peer-to-peer public key infrastructure that utilizes a ``web of trust" to add another layer of transparency and decentralization to the model proposed in this paper. There is also ongoing research in the area of implementing peer-to-peer public key infrastructure in blockchains such as Ethereum by utilizing smart contracts to create and authenticate identities in a distributed manner. If this type of authentication could be built into the chaincode of Hyperledger Fabric and utilized for the framework presented in this paper it could create an easier system of authentication that relies less on centralization. Other improvements to the framework could be made such as additional chaincode functionality (e.g., meeting windows, reputation system, designated driving zones).

%%%%%%%%%%%%%%%%%%%%%
%\input{FutureWork.tex}

\bibliographystyle{IEEEtran}
\bibliography{References}

%%%%%%%%%%%%%%%%%%%%%%%%%%%%%%%%%%%%%%%%%%%%%%%%%%
\vspace{-30pt}
\begin{IEEEbiography}[{\includegraphics[width=1in,height=1.25in,clip]{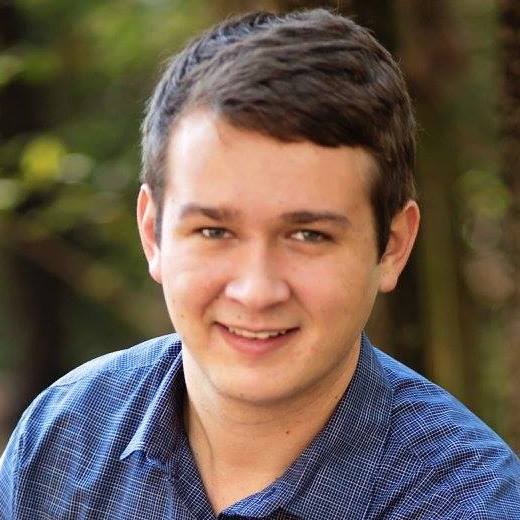}}] {Ryan Shivers} is pursuing his MS degree in Computer Science at Tennessee Technological University, USA. He received his BS in Computer Science with a concentration in Scientific and Software Applications from Tennessee Technological University in 2017. Ryan’s primary research interests are in application of secure blockchain architectures to new domains, security of information flowing through blockchain smart contracts, malware analysis, and security automation tooling.  Ryan has experience working with securing both Ethereum and Hyperledger Fabric smart contracts and has performed research on utilizing machine learning algorithms to identify malicious smart contract invocation transactions. Outside of research Ryan works on developing penetration testing skills by participating in capture the flag competitions.
\end{IEEEbiography}
%\vfill

\vspace{-30pt}
\begin{IEEEbiography}[{\includegraphics[width=1in,height=1.25in,clip]{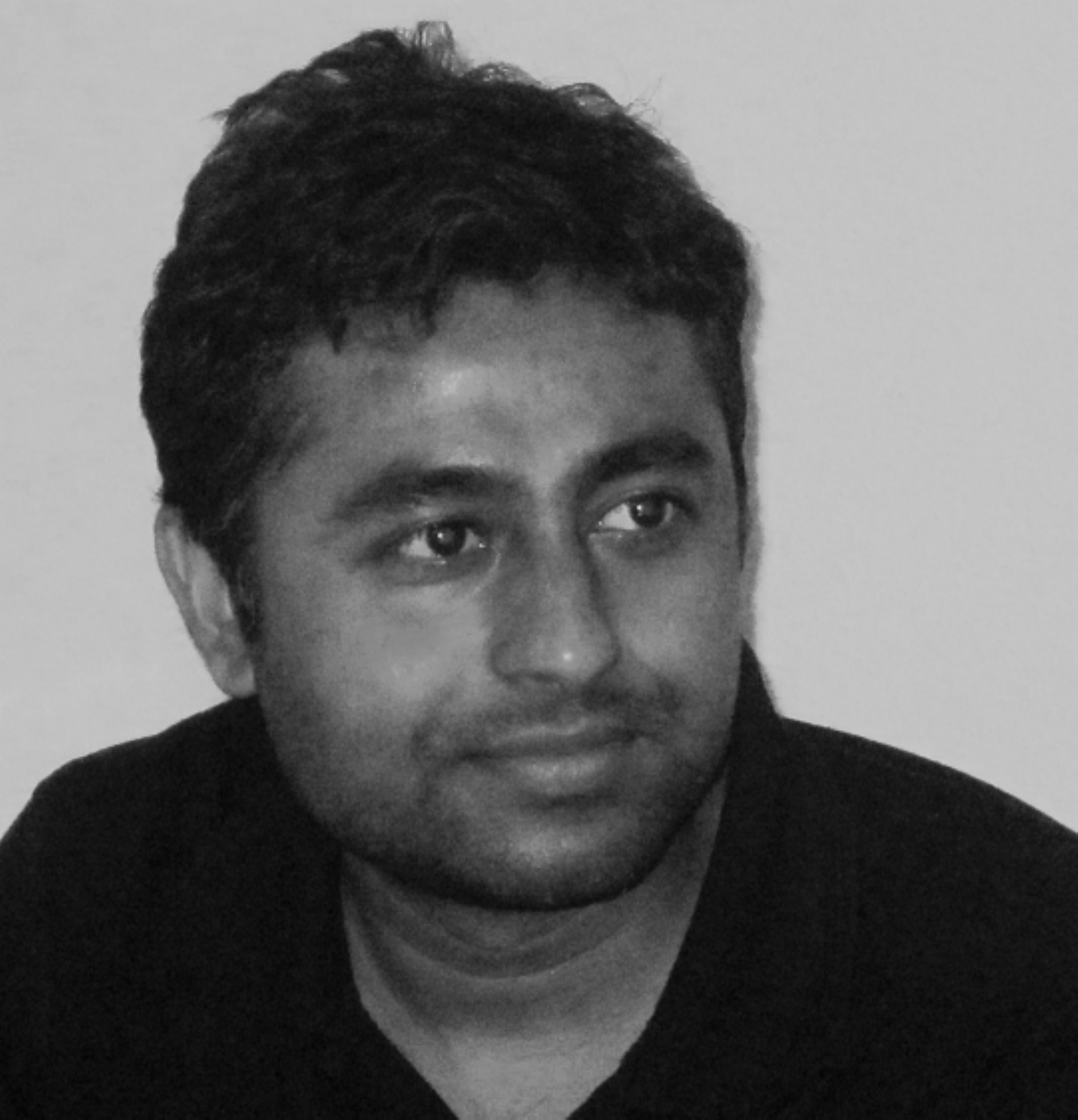}}] {Mohammad Ashiqur Rahman} is an Assistant Professor in the Department of Electrical and Computer Engineering at Florida International University (FIU), USA. Before joining FIU, he was an Assistant Professor at Tennessee Tech University. He received the BS and MS degrees in computer science and engineering from Bangladesh University of Engineering and Technology (BUET), Dhaka, in 2004 and 2007, respectively, and obtained the PhD degree in computing and information systems from the University of North Carolina at Charlotte in 2015. 
Rahman's primary research interest covers a wide area of computer networks and communications, within both cyber and cyber-physical systems. His research focus primarily includes computer and information security.
%, risk analysis and security hardening, secure and dependable resource allocation and optimal management, and distributed and parallel computing. 
He has already published over 50 peer-reviewed journals and conference papers.  He has also served as a member in the technical program and organization committees for various IEEE and ACM conferences.
\end{IEEEbiography}

\vspace{-30pt}
\begin{IEEEbiography}[{\includegraphics[width=1in,height=1.25in,clip]{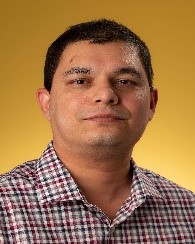}}] {Hossain Shahriar} is an Associate Professor of Information Technology and BSIT/BASIT Program Coordinator at Kennesaw State University, Georgia, USA. His research interests include mobile and web security, EMR and healthcare security, malware analysis. One of his research areas is focusing on automatic checking for vulnerabilities in implementation and design of applications, development of tools towards development of educational resources and capacity building on Secure Mobile Software Development (SMSD, https://sites.google.com/site/smsdproject/home), a project sponsored by National Science Foundation. He is currently investigating healthcare data interoperability issues and potential solutions, particularly adopting smart conract-based distributed apps. His research works also spanned over clinical workflow analysis, process mining, securing of Electronic Health Record Systems, smart and connected health tool. Dr. Shahriar has published over 85 peer-reviewed articles in IEEE/ACM conferences, journals and book chapters including ACM Computing Surveys, Computers and Security, and Future Generation Computer Systems.
\end{IEEEbiography}

\end{document}